# Does high harmonic generation conserve angular momentum?

Avner Fleischer, Ofer Kfir, Tzvi Diskin, Pavel Sidorenko, and Oren Cohen

Solid state institute and physics department, Technion, Haifa, 32000, Israel
emails: avnerf@tx.technion.ac.il,   oren@technion.ac.il

**Abstract**

High harmonic generation (HHG) is a unique and useful process in which infrared or visible radiation is frequency up-converted into the extreme ultraviolet and x-ray spectral regions. As a parametric process, high harmonic generation should conserve the radiation energy, momentum and angular momentum. Indeed, conservation of energy and momentum have been demonstrated. Angular momentum of optical beams can be divided into two components: orbital and spin (polarization). Orbital angular momentum is assumed to be conserved and recently observed deviations were attributed to propagation effects. On the other hand, conservation of spin angular momentum has thus far never been studied, neither experimentally nor theoretically. Here, we present the first study on the role of spin angular momentum in extreme nonlinear optics by experimentally generating high harmonics of bi-chromatic elliptically-polarized pump beams that interact with isotropic media. While observing that the selection rules qualitatively correspond to spin conservation, we unequivocally find that the process of converting pump photons into a single high-energy photon does not conserve angular momentum, i.e. this process is not self-contained. In one regime, we numerically find that this major fundamental discrepancy can be explained if the harmonic photons are emitted in pairs. Yet in another regime, the ionizing electron carries the missing angular momentum. The results presented here, apart from exploring the very foundations of HHG, are also important for a variety of applications – as our system exhibits full control over the harmonics polarization, from circular through elliptical to linear polarization, without comprising the efficiency of the process. This work paves the way for a broad range of applications with HHG, from ultrafast circular dichroism to zepto-clocks and to attosecond quantum optics.

# Introduction

High harmonic generation (HHG) has been providing unique experimental access to ultrafast processes in atoms [1], molecules [2,3], solids [4] and plasmas [5] as well as for high resolution imaging [6] and spectroscopy of heat transfer in the nano-scale [7]. HHG from gases results from laser-induced recollision between a parent ion and an electronic wavepacket, following its release into the continuum by tunnel ionization and subsequent evolution under the influence of the driver field [8]. The short-wavelength photon is emitted when the unbounded highly-energetic electronic wave-packet recombines into the bound state from which it had tunneled out. For this reason, HHG from gas media is considered a parametric process where the atom completely returns to its original quantum state and a single high-energy photon is emitted.

Parametric pertubative processes in isotropic media, such as third harmonic generation and sum frequency generation, are described as exchange of photons [9]. Accordingly, the energy, momentum and angular momentum of the emitted photon correspond to the net energy, momentum, and angular momentum of the (annihilated) pump photons, respectively [10]. The situation is less clear in non-perturbative parametric processes such as HHG [11]. While a photon-exchange model for HHG has not been derived from first principles, all current HHG experiments conform to conservations of energy and momentum. The conservation of energy was demonstrated, for example, by high-order wave-mixing of bichromatic drivers [12], and the conservation of momentum was nicely demonstrated in a non-collinear configuration [13]. With angular momentum the situation is completely different. Angular momentum of paraxial beams can be divided to orbital and spin components, which are associated with beam "vorticity" and polarization, respectively. Recently, generation of high-order harmonics with orbital angular momentum was reported [14], and the observed deviations from orbital angular momentum conservation were attributed to noise and propagation effects. **Here, we present the first experimental and theoretical study on the conservation of spin angular momentum in HHG. Our experiments study the effects at the single atom level, with additional motivation to control the polarization of the high harmonics.**

Symmetry arguments dictate that the polarization of high harmonics produced during interaction between linearly-polarized drivers and isotropic media will be linear [15]. This is indeed the case in the vast majority of HHG studies and applications. Two methods have been experimentally demonstrated for generation of elliptically-polarized high-order harmonics, yet they both displayed relatively small ellipticity (<0.37): irradiating atoms with an elliptically-polarized driver [16,17] and aligned molecules with a linearly-polarized driver [18,19]. These methods rely on non-"head-on" recollisions that also significantly decrease the HHG yield inevitably [20]. In addition, the controllability over the harmonics polarization in those methods is poor due to their complicated non-linear dependence on the experimental knobs. Likewise, circularly polarized high-order harmonics were produced by phase-shifting the polarization of linear harmonics using a reflective quarter-waveplate [21]. This method is unfortunately very lossy. Thus, experimental control over the polarization state of HHG radiation is still very limited to this day. Nevertheless, there is strong motivation for extending the range of accessible polarization-states in HHG, as polarization is a fundamental property of light and of light-matter interactions. Indeed, many theoretical methods were proposed for generation of circularly-polarized (e.g. [22,23]) and elliptically-polarized (e.g. [24,25]) high harmonics. Especially notable is a method suggested for generation of circularly-polarized high harmonics by irradiating atoms with a coplanar, circularly-polarized counter-rotating bi-chromatic drivers [26]. A single pioneering experimental work from 1995 [27] reported the generation of high harmonics using this geometry. While the correct selection rules were obtained in that experiment, no attempt was made for measuring the polarization state of the harmonics.

Here we propose and demonstrate a simple method for generating high-order harmonics with fully controlled polarization, from linear through elliptical to circular polarization. Importantly, the conversion efficiency to harmonics with arbitrary polarization is comparable to that of the standard HHG process yielding linearly-polarized high harmonics driven by a linearly polarized laser pulse (see Figures S.1 and S.2 in Supplementary Information). Thus, the new procedure described here opens the door to numerous applications of HHG, where fully controllable polarization in the extreme UV and x-ray regimes offer new tools for imaging, spectroscopy, and more. Our approach is based on wave-mixing of two bi-chromatic drivers with controlled spin angular momentum in isotropic medium (gas). The drivers co-propagate and their polarization is elliptic with opposite helicity. We specifically experiment with 800nm (Ti:Sapphire) driver and its 1.95 harmonic. This wave-mixing of incommensurate frequencies allows us identifying the number of photons that each driver contributes to the generation of every harmonic-order. Varying the ellipticities of the driver beams leads to rich selection rules that qualitatively correspond to conservation of spin angular momentum. However, our experiments unequivocally show that the process of converting pump photons into a single high-energy photon does not conserve spin angular momentum, i.e. this process is not self-contained. We find numerically that this discrepancy is resolved differently in two different regimes. When the 410 nm pump beam is circularly polarized and the 800 nm pump beam is elliptically polarized, the spin angular momentum is not conserved for generation of each high harmonic separately, but it is conserved if the harmonics photons are emitted in pairs. In another regime, when the 800 nm pump beam is circularly polarized and the 410 nm beam is elliptically polarized, the radiation alone cannot balance the angular momentum. In this case, the missing angular momentum is imparted into the ionizing electron, making the HHG process a non-parametric one.

**Experimental setup**

To explore the role of spin angular momentum (SAM) in extreme nonlinear optics and control the polarization of the high harmonics, we mix co-propagating elliptically-polarized waves in isotropic media. In our setup, sketched in Fig. 1a, a 1KHz repetition-rate Ti:Sapphire laser delivers linearly-polarized 40fs, 2mJ pulses, centered around 800nm with a bandwidth of 60nm. This beam is focused into a 100 micron Type I BBO crystal, where a second harmonic field is generated. By slightly tilting the BBO crystal, we red-shift the spectrum of the second harmonic beam such that it is centered around 410 nm. The two co-propagating colors are then separated in a Mach-Zehnder-like Interferometer (where dichroic mirrors are used as beam splitters and combiners), and the polarization state of the light in each arm is controlled separately by achromatic zero-order quarter-waveplates. The two beams are recombined and focused into a supersonic jet of Argon gas, yielding similar intensities at the focus ($\sim 2 \cdot 10^{14}$ W/cm$^2$). The emitted high harmonics are analyzed in a HHG spectrometer and their polarization is measured by an XUV reflective polarizer consisting of 3 fused silica blanks. The polarizer favors the transmission of *s*- polarized high harmonics with extinction ratio of 40:1 around harmonic 19. As explained below, the wave-mixing between such beams with an incommensurate frequency ratio ($\omega$ and $1.95\omega$) allows us to identify the number and color of absorbed pump photons in the process of generating high-order harmonics.

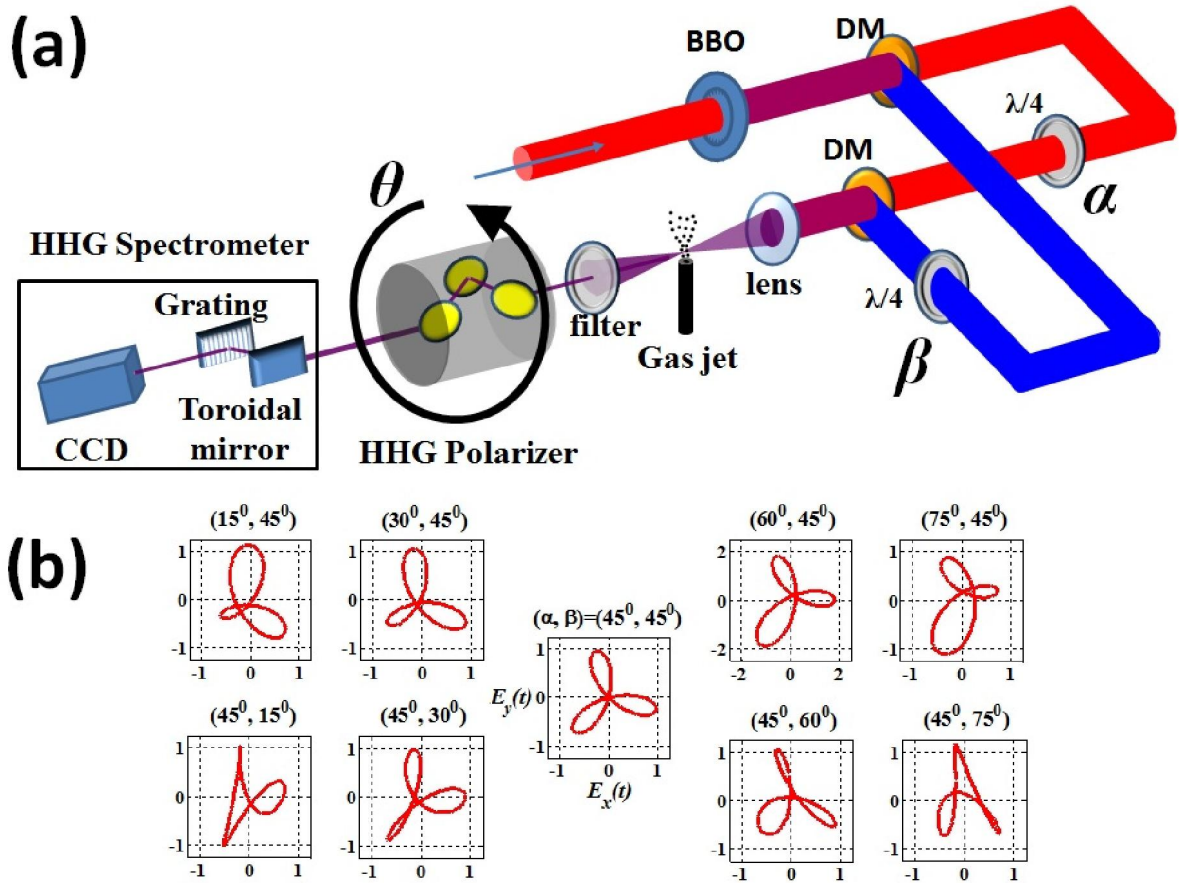

Figure 1: (a) Setup layout showing independent polarization control of the two-color drivers, the HHG polarizer and the HHG spectrometer. DM-dichroic mirrors; (b) Lissajous curves of the total electric field experienced by the Argon atoms in the jet for several readings of the quarter-waveplates. When both waveplates are set at $45^0$, the two-colors are counter-rotating circularly polarized, and the total electric field has a rosette-like shape having a 2.95-fold rotational symmetry. As the reading of either of the waveplates deviates from $45^0$, this symmetry is lost.

**Spin angular momentum in HHG**

Figure 2 shows some experimental and numerical high harmonics signals vs. harmonic order and the readings on one of the quarter-waveplates ($\alpha$ or $\beta$). For the sake of simplicity, the orientation of only single waveplate is varied at a time, and the second is held fixed at $45^0$, so as to yield a circularly-polarized field at that color. Figure 2a shows the experimental scan for which $\alpha$ is scanned and $\beta=45^0$, i.e., the second harmonic beam is always right-handed (clockwise) circularly polarized. Figure 2b shows the same scan in logarithmic scale. Due to the necessary high spectral resolution of the HHG spectrometer (which is inevitably accompanied by a small field of view), only the 18th-21st harmonics are shown in the trace. This does not restrict the generality of our measurements. Figure 2c shows a numerical scan obtained under similar conditions. The numerical scan is obtained by calculating the dipole acceleration expectation value obtained from a 3D time-dependent Schrodinger equation (3D TDSE) simulation of a single electron in a model of an Argon atom, interacting with a pulse of right-circularly-polarized 410nm driver and a left-elliptically-polarized 800nm driver. Figures 2d,e show experimental scans for which β is scanned and α=45$^0$ (left-handed circularly-polarized 800nm light), and Figure 2f shows a numerical scan obtained under similar conditions. We emphasize that the high harmonics in our scheme are produced with comparable intensity to the harmonics emitted when using bichromatic linear drivers (see Figures S.1 and S.2 in the Supplementary Information).

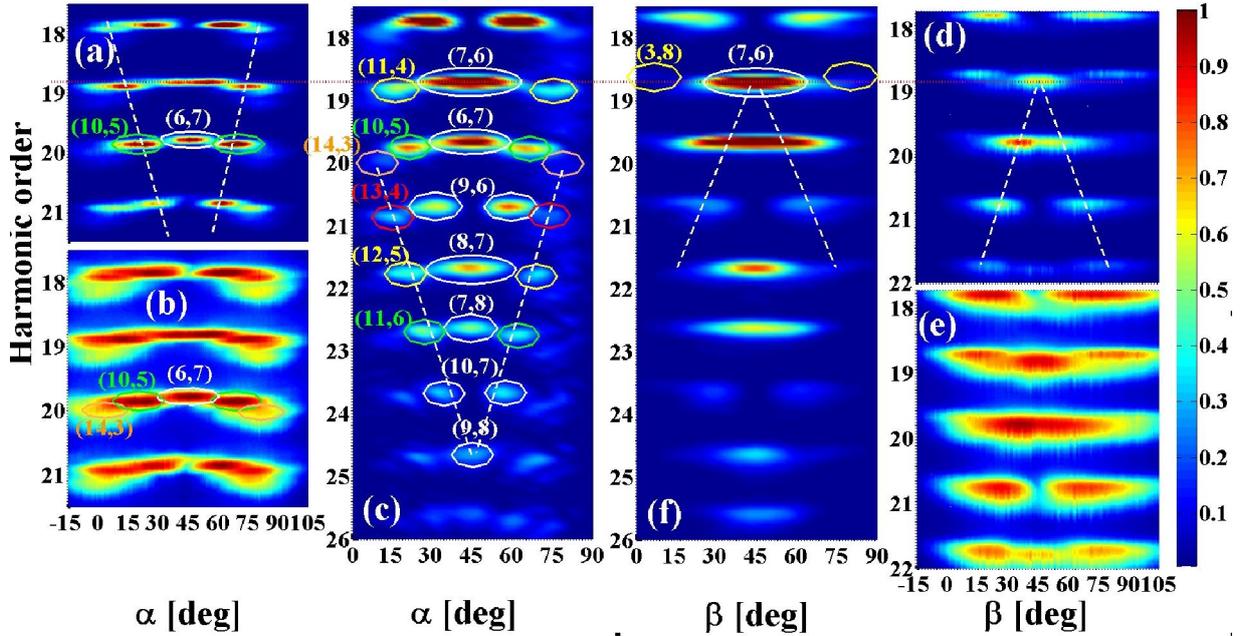

Figure 2: (a) Experimental HHG spectra vs. harmonic order and the angle α of the 800nm quarter waveplate. The 410nm source is held right-circularly-polarized. The harmonic appearing at the order 19.65 of the fundamental is labeled by a pair of two integer numbers (6,7) corresponding to absorption of 6 red photons and 7 blue photons. As the waveplate reading deviates from 45$^0$ an additional channel (10,5) appears. (b)- same as (a), shown in logarithmic scale, which better resolves the appearance of even the higher-order channel (14,3). (c)- HHG spectra obtained from numerical simulation under the same conditions as in (a). As before, the identified high-harmonic channels, which appear at harmonic orders $\Omega=n_1+1.95n_2$, are specified by the integer pair ($n_1$,$n_2$). (d), (e), (f)- same as (a), (b), (c), respectively, but where the angle β of the 410nm quarter waveplate is scanned and the 800nm source is held left-circularly-polarized.

Before discussing the role of spin angular momentum in the spectra displayed in Fig. 2, we discuss the effect of energy and parity conservation. From energy conservation, the harmonic frequencies in our bi-chromatic HHG experiment are given by:

$$\Omega_{(n_1,n_2)} = n_1 \cdot \omega + n_2 \cdot 1.95\omega \qquad (1)$$

where $n_1$ and $n_2$ are integer numbers that can be associated with the number of "driver photons" annihilated in the process from the drivers at angular frequencies $\omega$ and $1.95\omega$, respectively. Parity conservation requires that $n_1+n_2$ is odd. We can associate each harmonic in Fig. 2 with a unique pair $(n_1,n_2)$ because (1) other channels that yield the same harmonic frequency as $(n_1,n_2)$ [e.g. $(n_1+39,n_2-20)$] involve too many photons and (2) the spectral resolution of our HHG spectrometer is very high ($<<0.05\omega$). The identified channels $(n_1,n_2)$ are encircled and labeled in Fig. 2a,b,c,f. We note that our identification method cannot work when the two drivers have an integer frequency ratio. For example, channels $(n_1,n_2)$, $(n_1+4,n_2-2)$ and $(n_1-4,n_2+2)$ are all probable, and all yield the same harmonic frequency if the driver frequencies are centered at $\omega$ and $2\omega$.

As shown in Fig. 2, harmonics are born and vanish when the quarter-wave plates are rotated. We now show that conservation of angular momentum plays an important role in these selection rules. Notably, no theory describing transfer of angular momentum in HHG, at the single atom level, has ever been formulated yet. Thus, we propose the following simple model arising from our experiments. Adding an "extreme nonlinear optics correction term", $\delta_{(n_1,n_2)}$, to the perturbative form for conservation of spin angular momentum in the wave-mixing processes and get:

$$\sigma_{(n_1,n_2)} = n_1\sigma_1 + n_2\sigma_2 + \delta_{(n_1,n_2)} \qquad (2)$$

In Eq. (2), $\sigma_{(n_1,n_2)}, \sigma_1, \sigma_2$ are the spin expectation values (in units of $\hbar$) of the emitted HHG photon of channel $(n_1,n_2)$, and of the first and second driver, respectively. Within pertubative nonlinear optics, Eq. (2) with $\delta_{(n_1,n_2)} = 0$ is a direct consequence of Eq. (1) and manifests conservation of angular momentum, given that in our experiment SAM is the only sort of angular momentum present. However, all SAM expectation values have a very stringent constraint: they are bounded between -1 and 1, where 1 (-1) represent left (right) circular polarization. We start by comparing the measured and calculated spectra with the spectrum predicted in the perturbative limit of Eq. (2) (assuming $\delta_{(n_1,n_2)} = 0$). We begin with the counter-rotating circular driver case ($\alpha=\beta=45^0$). Since $\sigma_1 = 1, \sigma_2 = -1$ and $-1 \leq \sigma_{(n_1,n_2)} \leq 1$, it turns out that $n_1$ and $n_2$ must differ by unity, i.e., the allowed channels are only $(n_1,n_1\pm1)$. This condition is clearly satisfied by the experimental data displayed in Fig. 2. As we rotate the wave plate of the 800 nm driver, we change the beam ellipticity, $\varepsilon_1$, and its SAM according to $\sigma_1 = 2\varepsilon_1/(\varepsilon_1^2+1) = 2\tan\alpha/(\tan^2\alpha+1) = \sin(2\alpha)$ [See supplementary information]. Substituting this relation and the condition $|\sigma_{(n_1,n_2)}| \leq 1$ in the perturbative version of Eq. (2) assigns an allowed "existence region" for each channel $(n_1,n_2)$: $\sin^{-1}[(n_2-1)/n_1]/2 \leq \alpha \leq \sin^{-1}[(n_2+1)/n_1]/2$. These predicted existence regions conform qualitatively with many observed (Figs. 2a,b) and calculated (Fig. 2c) spectral features. It also explains why $n_1 \geq n_2 - 1$ in all the channels. For channel (7,6) for instance, the model predicts existence in the regime $22.8^0 \leq \alpha \leq 67.2^0$, in accordance with the experimental and numerical results. It also explains why the channels (13,4), (12,5), (11,6), (10,7), (9,8) in Figures 2a-c appear at values closer and closer to α=45⁰, giving rise to the "v-shape" seen as a white dashed lines connecting these harmonic channels in Figures 2a,c. A similar calculation explains the "opposite v-shape" seen in Figures 2d,f when β is varied: in this case, the existence regime of each channel reads: $\sin^{-1}\left[(n_1-1)/n_2\right]/2 \leq \beta \leq \sin^{-1}\left[(n_1+1)/n_2\right]/2$. All these correspondences indicate that conservation of angular momentum plays an important role in determining the observed selection rules and that a simple model based on perturbative nonlinear optics gives qualitatively good predictions for our extreme nonlinear optics experiment.

**Controlling the polarization of HHG**

Next, we show that the polarization of high harmonics in our scheme is fully controlled by rotating one of the quarter-wave plates. We measure the polarization of several harmonics by rotating our extreme ultraviolet polarizer (some transmission traces versus polarizer angle are presented in Fig. S.4 in the Supplementary Information section). We numerically calculate the polarization of the high harmonics using our 3D TDSE solver. In addition, we calculate the predicted polarization assuming the perturbative model in Eq. 2. For fixing the wave-plate of the 410nm driver at β=45⁰, yielding right circular polarization for the that driver, we substitute $\delta_{(n_1,n_2)}=0$, $\sigma_2=-1$, $\sigma_1=\sin(2\alpha)$ and $\sigma_{(n_1,n_2)}=2(h\cdot\varepsilon)_{(n_1,n_2)}/\left(\varepsilon^2_{(n_1,n_2)}+1\right)$ to Eq. (2) and get by simple algebra that: $(h\cdot\varepsilon)_{(n_1,n_2)}=\left(n_1\sin(2\alpha)-n_2\right)^{-1}\pm\sqrt{\left[1-\left(n_1\sin(2\alpha)-n_2\right)^2\right]/n_1\sin(2\alpha)-n_2}$ where $\varepsilon_{(n_1,n_2)}$ is the ellipticity of the (n₁,n₂) harmonic channel and $h_{(n_1,n_2)}$ is its helicity. A similar expression is obtained for the case of α=45⁰, where the 800nm field is left circularly polarized and the 410nm driver ellipticity is scanned.

The calculated and measured ellipticities of the high harmonics are displayed in Fig. 3. Figure 3(a-c) show results for the case of a fixed waveplate of the 410nm driver. The calculated and measured polarization (product of ellipticity and helicity) and intensity of the (7,6) channel are shown in Fig. 3a and Fig. 3b, respectively. Four important results are presented in these plots. First, circular (ε=0.95±0.09) and elliptic (ε=0.7±0.09) polarization of the channel (7,6) are experimentally demonstrated, significantly extending the current state-of-the-art ellipticitiy-range of HHG. Second, the ellipticity of the harmonic is fully controlled by rotating the waveplate of the 800nm driver. According to the numerical calculation, for α=22.8⁰ we get $\varepsilon_{(7,6)}(\alpha=22.8^0)=0$ (linearly-polarized light). As α increases to α=29.5⁰, 36.5⁰, 45⁰, and 54.5⁰ the ellipticity of the harmonics changes all the way to unity (left-circular light), zero, unity (right-circular light), and zero again. Third, while the harmonic polarization varies from circular through elliptic to linear, its intensity varies by less than 20%. Fourth, the prediction based on the perturbative model matches the numerical calculation and experimental measurements only in the region of α~45⁰ (Fig 3b), indicating that the strong-field correction becomes significant as the wave-plate is rotated away from α~45⁰. Figure 3c shows the numerical calculation for the ellipticity-helicity product for a wide spectral region versus the wave plate angle. It shows that the polarization behavior displayed in Fig 3a for the (7,6) channel is general. Interestingly, the (n₁,n₁-1) and (n₁,n₁+1) groups of channels exhibit mirror-like polarization dynamics (see also red and green traces in Fig 3b) . Finally, we show in Figs 3(d-f) that the polarization of the high harmonics can also be controlled by rotating the wave-plate of the 410 driver. Interestingly, the perturbative model matches the numerical calculation for a substantial range of wave-plate angles (see the correspondence in Fig. 3e). A time-dependent perspective for the origin of polarization control, which is based on the recollision phenomenon, is presented in the supplementary material.

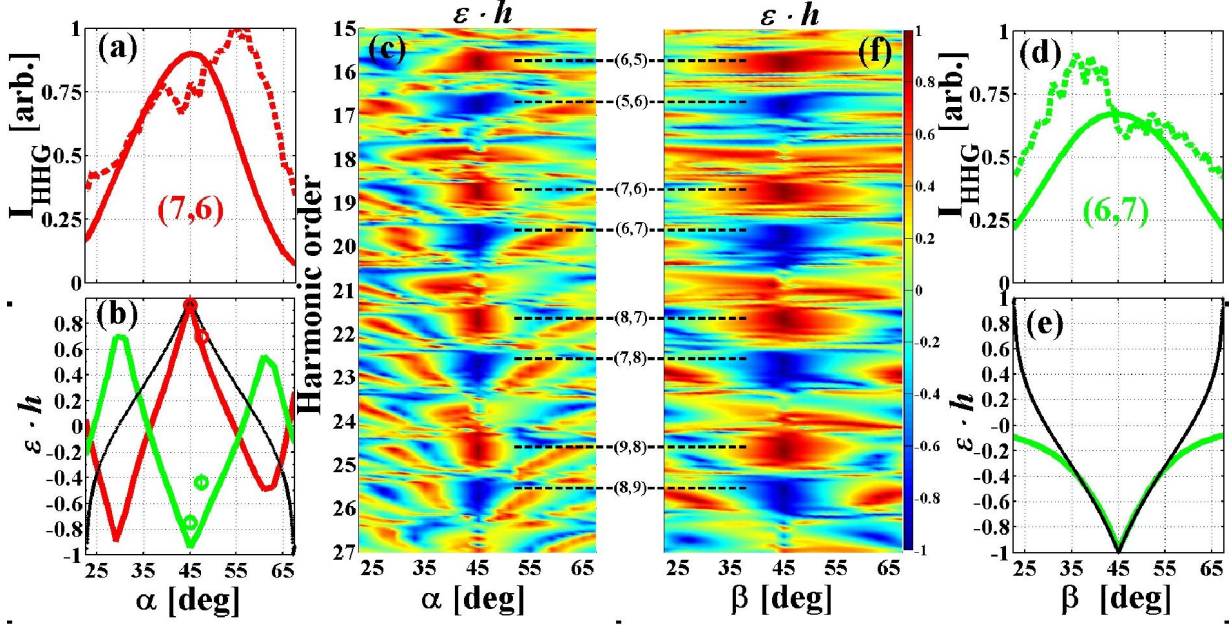

Figure 3: (a)- Numerical (solid red curve) and measured (dashed red curve) HHG spectra of the channels (7,6). (c) Numerical ellipticity times helicity of the polarization ellipse of the emitted harmonics vs. harmonic order and the angle α of the 800nm quarter waveplate. The 410nm source is held right-circularly-polarized. (b)- Numerical ellipticity times helicity 2D traces taken from (c) for the channel (7,6) (red curve), as well as the theoretical dependence predicted by Eq. 2 when $\delta_{(n_1,n_2)}$ is neglected (black curve). Also shown are two measured values (red circles). The ellipticity of harmonics ($n_2\pm1$, $n_2$) could be controlled from perfect right-circularly-polarized through linear and then to perfect left-circularly-polarized as α is canned from $45^0$ to about $27^0$. "mirror-like" result are demonstrated for the channel (6,7) (green curve and circles) (d), (e), (f)- same as (a), (b), (c), respectively, but for the channel (6,7) (green curve) where the angle β of the 410nm quarter waveplate is canned and the 800nm source is held left-circularly-polarized. The matching between the theoretical prediction for the ellipticity [black curve in (e)] and the numerical calculation (green curve) applies for a larger interval of angles β.

**Coupled high-energy photons**

In previous sections we showed that many observed and calculated results qualitatively conform to conservation of energy and spin angular momentum in a process of exchanging multiple photons from the drivers into a single high-energy photon (i.e., Eq. (1) and Eq. (2) with $\delta_{(n_1,n_2)} = 0$). We now focus on the deviations from this correspondence and then discuss their implications. Such deviations mean that in some regime angular momentum is not conserved – for the basic process (which is believed to be the only dominant process of HHG) of converting multiple pump photons into a single high harmonic photon. The deviation is most pronounced for the ($n_1$, $n_1+1$) channels in the α-scan. For example, consider channel (6,7) with $\beta=45^0$ ($\sigma_2=-1$). Conservation of spin angular momentum implies that (1) the spin angular momentum and ellipticity-helicity product can be only -1 and that (2) this channel should not exist for $\alpha\neq45^0$. These predictions are in sharp contrast with both the numerical and experimental results (green curve and circles) in Fig. 3b. As a second example, consider channel (11,6) and compare its predicted and actual "existence regions" (Fig. S.3 in Supplementary Information). The perturbative model predicts existence in the regime $13.5^0 \leq \alpha \leq 19.7^0$, while in the experiments this channel appears in the regime $22^0 \leq \alpha \leq 30^0$. On the other hand, the prediction of Eq. (2) with $\delta_{(11,6)} \cong -3$ agrees with the observations. In general, for most harmonic channels, it seems that the strong-field correction $\delta_{(n_1,n_2)}$ becomes significant as the wave-plate is rotated away from $\alpha \sim 45^0$ (see Supplementary Information).

We note that in Figures 3b,c, the $(n_1,n_1-1)$ and $(n_1,n_1+1)$ groups of channels exhibit mirror-like polarization dynamics. Moreover, as α is scanned from below $45^0$, through $45^0$ and above $45^0$, these groups exchange the orientations of their polarization ellipses (Figure S.5 in the Supplementary Information). This suggests the possibility of correlation between these two groups of harmonics, which transfer the missing SAM from one to the other. This can happen only if two photons at different harmonics are emitted correlatively, and only the total SAM is conserved.

In order to test this hypothesis, we plot in Fig. 4 the strong-field corrections $\delta_{(n_1,n_2)} = 2(h \cdot \varepsilon)_{(n_1,n_2)} / (\varepsilon^2_{(n_1,n_2)} + 1) - n_1 \sin(2\alpha) + n_2$ where $\varepsilon_{(n_1,n_2)}$ is the numerically calculated ellipticity. We clearly see that these corrections reach very large values. This means that, for each harmonic channel alone, energy conservation is not accompanied by spin conservation (Eq. 2 in the perturbative limit). However, as shown in Fig. 4a, the averaged sum $\bar{\delta} = \sum_{(n_1,n_2)} N_{(n_1,n_2)} \cdot \delta_{(n_1,n_2)} / \sum_{(n_1,n_2)} N_{(n_1,n_2)}$ is close to zero. Here, $N_{(n_1,n_2)}$, a measure for the number of photons emitted into the $(n_1,n_2)$ channel, is the peak intensity of that channel divided by its frequency. That is, in the process of transforming the driver photons into the high harmonic ones, the conservation of energy-spin is fulfilled not for each channel alone, but rather for the entire emission. Hence, the radiative process of HHG is indeed parametric, but in this case only because the high harmonics are coupled. Interestingly, intensity pairing between high harmonics of different orders driven by bi-chromatic drivers, was observed in Ref. [28]. The nature of correlation could be further investigated by calculation of the partial sums $\delta_{(n_1-1,n_1)} + \delta_{(n_1+1,n_1)}$ (rather than $\bar{\delta}$) which turn out to be close to zero in some range of α. This shows that to some extent the energy-spin conservation is maintained even within channel-pairs, for instance channels (5,6) and (7,6). This suggests that the atom converts 12 photons from each driver into two coupled high-harmonic photons.

The parameter $\bar{\delta}$ evaluates the difference between the spin angular momenta of the absorbed and emitted radiation. Thus, $\bar{\delta} \neq 0$ indicates that the processes is not parametric. In Fig. 4 we show that $\bar{\delta} \approx 0$ for all angles of the 800 nm wave-plate which means that the process is parametric. We now show that the process can become nonparametric if the other wave-plate is rotated. Figure 5 shows the strong-field corrections as a function of β: $\delta_{(n_1,n_2)} = 2(h \cdot \varepsilon)_{(n_1,n_2)} / (\varepsilon^2_{(n_1,n_2)} + 1) - n_1 + n_2 \sin(2\beta)$. The strong-field corrections are close to zero in some range of β and are positive elsewhere. It is obvious that no partial sums could remove these large deviations from zero. Even the full averaged sum turns out to be larger than it was in the α-scan. The only physical entity that can carry the missing angular momentum in our simulations is the electron. Since the bound part of the electron returns to its initial ground state after the driver fields have subsided, we conclude that, in this case, it is the ionizing electron which carries the missing angular momentum.

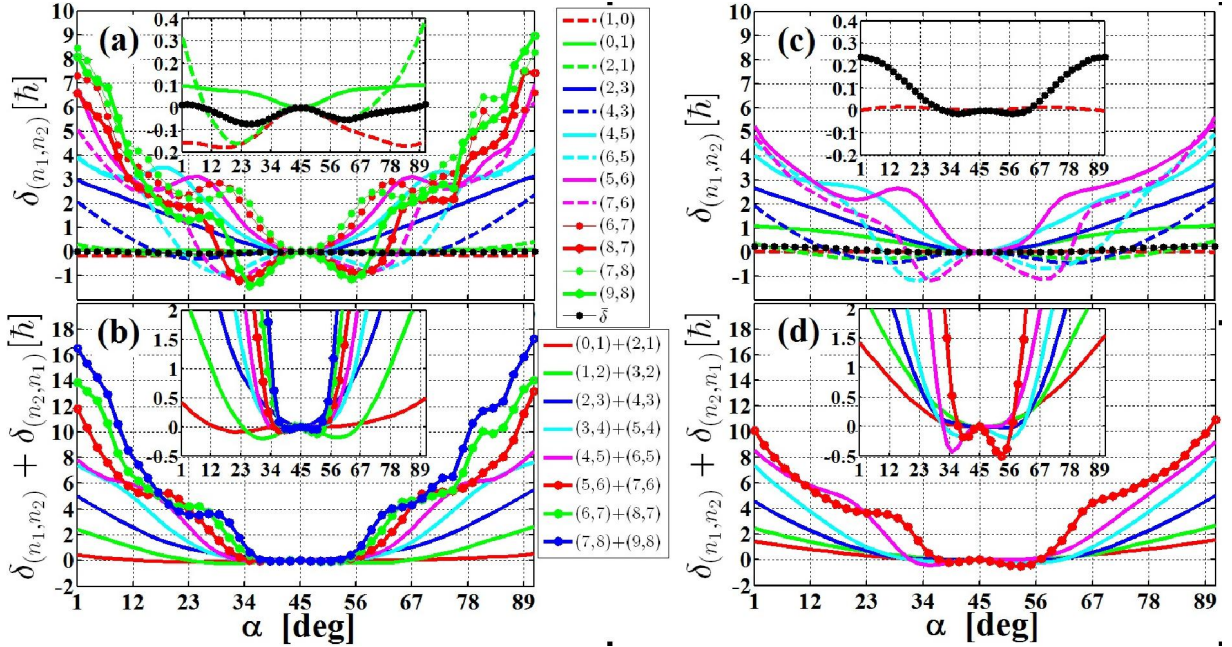

Figure 4: (a)- Numerical $\delta_{(n_1,n_2)}$ values calculated from $\delta_{(n_1,n_2)} = 2(h \cdot \varepsilon)_{(n_1,n_2)} / (\varepsilon^2_{(n_1,n_2)} + 1) - n_1 \sin(2\alpha) + n_2$ for different channels ($n_1,n_2$). $\bar{\delta}$ is the weighted sum over all existing channels in the harmonics spectrum. While for each harmonic channel alone $\delta_{(n_1,n_2)}$ differs from zero a lot, the weighted sum is indeed close to zero, indicating that in the process of transforming the driver photons into the high harmonic ones, the conservation of energy-spin is fulfilled not for each channel alone, but rather for the entire emission. Hence, the radiative process of HHG is parametric, and the high harmonics are correlated. (b) The partial sums $\delta_{(n_1-1,n_1)} + \delta_{(n_1+1,n_1)}$ which are also close to zero in some range of α, show that to some extent the energy-spin conservation is maintained even within channel-pairs, for instance channels (5,6) and (7,6). This suggests that the atom converts 12 photons from each driver into two coupled high-harmonic photons. (c),(d)-same as (a),(b), but when the drivers intensities are reduced such that the total ionization fraction reduces from 0.8 [(a) and (b)] to 0.06 [(c) and (d)]. Since no much difference is observed in the large values of $\delta_{(n_1,n_2)}$, and since energy-spin conservation is again manifested through the small value of $\bar{\delta}$, despite of the fact that the ionization fraction has reduced by an order of magnitude, it is concluded that the ionizing electron doesn't carry any excess energy-spin, making the HHG process indeed parametric.

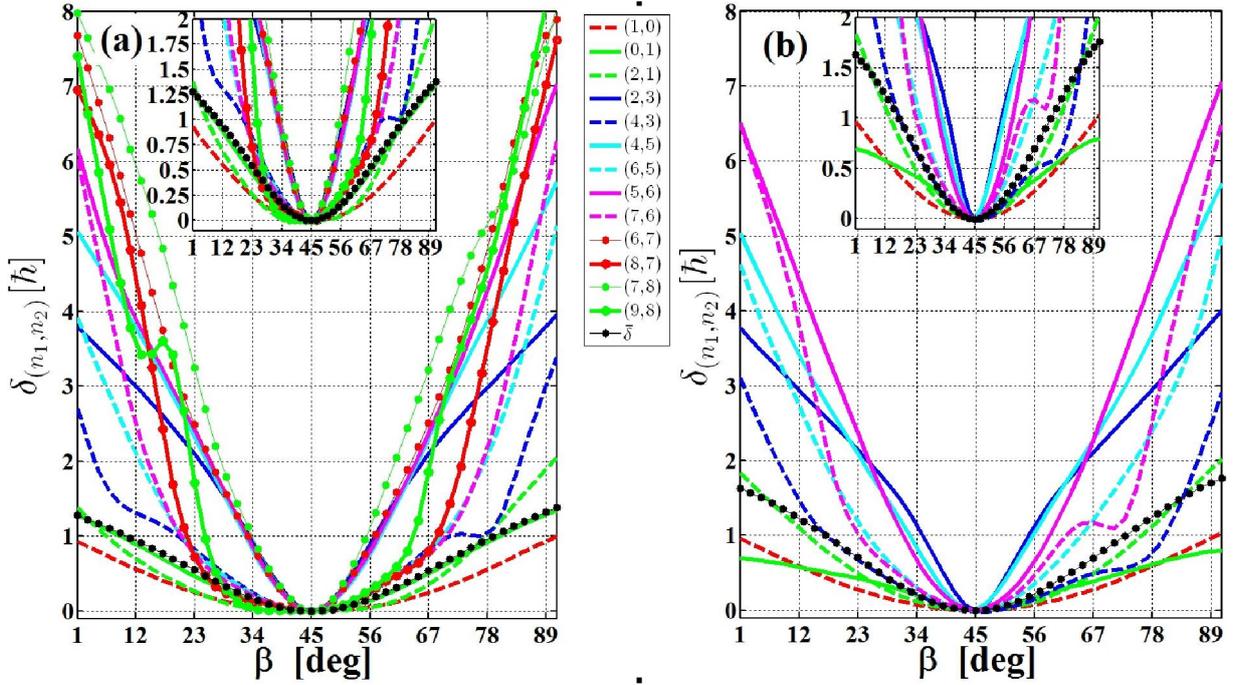

Figure 5: (a)- Numerical $\delta_{(n_1,n_2)}$ values calculated from $\delta_{(n_1,n_2)} = 2(h \cdot \varepsilon)_{(n_1,n_2)} / (\varepsilon^2_{(n_1,n_2)} + 1) - n_1 + n_2 \sin(2\beta)$ for different channels ($n_1$,$n_2$). Here, both the individual corrections $\delta_{(n_1,n_2)}$ and the weighted sum $\overline{\delta}$ differs from zero a lot, indicating that the seemingly violation of conservation of energy-spin is accompanied by the ionizing electron carrying the missing spin. Hence, the radiative process of HHG is non-parametric. (b)-same as (a), but when the drivers intensities are reduced such that the total ionization fraction reduces from 0.8 [(a)] to 0.06 [(b)]. Since not much difference is observed in the large values of $\delta_{(n_1,n_2)}$, $\overline{\delta}$, despite of the fact that the ionization fraction has reduced by an order of magnitude, it is concluded that the ionizing electron carries that same amount of excess spin as before.

Two different cases are presented in Fig.4 and Fig.5: one in which $\overline{\delta}$ is close to zero and another in which $\overline{\delta} \neq 0$. We argued that in the first case energy-spin conservation is maintained within the entire emission, while in the second case the ionizing electron must carry the excess angular momentum. Repeating the two simulations with reduced driver intensities (such that the ionization fraction is decreased from about 0.8 to 0.06) yield very similar values for $\delta_{(n_1,n_2)}$, indicating that the mechanisms by which energy-spin is conserved are inherent to the physical process. This fact has another important experimental consequence: the ellipticities of the different channels $\varepsilon_{(n_1,n_2)}$ do not depend on the driver intensities. Moreover, we confirmed that they are also independent from the intensity ratio between the drivers. This intensity-independence feature is very important because otherwise the harmonics field produced at a focus of the driver beams, where the intensity varies greatly in the transverse plane, would exhibit a transversely-varying polarization state, including varying ellipticity. But our experiments rule this option out.

**Conclusions and outlook**

This work paves the way for new research directions and numerous applications. We presented a simple and effective method to control the polarization of high harmonics, all the way from linear to circular. Our scheme is based on spin wave-mixing and as such it is general and can be implemented using broad range of laser systems, nonlinear media and spectral region [29]. Applications of this source can include ultrafast extreme UV and x-ray circular dichroism of magnetic films [30] molecules [31] and topological insulators [32], extreme UV zepto-clocks and laser-STM [33,34] and polarization-dependent absorption spectroscopy for exploring structural changes [35]. Imaging of magnetic domains with high spatio-temporal resolution [36] may lead to enhanced magnetization-switching rates [37,38]. In addition to downstream experiments, HHG spectroscopy using polarization measurements in combination with varying the ellipticities of bi-chromatic drivers may be used for tracking different HHG channels [39,40], attosecond time evolution of magnetic quantum numbers in matter, and probing circulation of electrons in molecules [41,42] or superconductors [43]. Moreover, the inclusion of photon spin degree of freedom to extreme nonlinear optics can lead to new features, including coupling between angular momentum of above threshold ionization and HHG. The observation that conservation of spin-energy quanta play a role in our experiment can provide a benchmark for a fully quantum theory of HHG [44]. Finally the numerical indication for coupling between high harmonics and high-energy photon pairs could lead to quantum optics with attosecond pulses.

*Note added:* Recently, a different technique was experimentally demonstrated for generation of elliptically polarized high harmonics. In that technique, the harmonics were driven by bi-chromatic drivers that are polarized linearly in orthogonal directions. Ellipticities up to 0.7 were reported. That work was presented in: G. Lambert et al., "Highly elliptical but intense high harmonics from gas in a two color field" P1.71. in the 4th international conference on attosecond physics, Paris, July 8-12, 2013. The work presented in this manuscript was shown one month before in QW1A.6, CLEO, Conference on Lasers and Electro-Optics, San Jose, CA, USA, 9-14 June 2013: A. Fleischer et al., "High-Order Harmonics of Bichromatic Counter-Rotating Elliptically-Polarized Drivers: Fully Controlled Polarization State and Novel Selection Rules".

# Supplementary Information

## 1. Time-dependent perspectives of the experiment

The selection rules given in Eq. 1 could be intuitively understood by analyzing the recollision process in the time domain. For the counter-rotating circular driver case ($\alpha=\beta=45^0$), substituting the integers ($n_1, n_1 \pm 1$) in Eq. (1) yields [28, 44-45]:

$$\Omega = (2.95 n_1 \pm 1) \omega \qquad (S1)$$

where $n_1$ is any integer. This equation reflects the 2.95-fold rotational symmetry that the total electric field draws in space [see the Lissajous curve (red line) in Figure S.1]. As usual, the electron would tunnel close to the maximum of the electric field and would recombine later on. In one optical cycle T of the driver $\omega$, 2.95 recollision events occur in the plane of polarization of the two drivers. These events occur T/2.95 one after another, along directions $2\pi/2.95$ ($122^0$) to each other, but in the limit of low depletion are otherwise identical. It is this time-space synchronization between the recollision events that leads to the selection rules given in Eq. (S1) upon the analysis of the time-dependent dipole acceleration expectation value. Suppose for simplicity that the first recollision burst $a_1(t)$ occurs along the positive x axis, then the time-dependent x-component of the total dipole acceleration in one optical cycle T would look like

$$a_x(t) = a_1(t) + \cos\left(\frac{2\pi}{2.95}\right) a_1\left(t - \frac{T}{2.95}\right) + \cos\left(\frac{4\pi}{2.95}\right) a_1\left(t - \frac{2T}{2.95}\right) \qquad (S2)$$

and in addition $a_x(t)$ is periodic in T. Hence, the only non-zero Fourier components of $a_x(t)$ turn out to be $2.95n \pm 1$. Repeating this analysis for the y-component of the total dipole acceleration would yield the same selection rules.

Recall that when HHG is driven by single color, linearly polarized laser, odd harmonics are obtained. Breaking the left-right symmetry (for instance, by adding a second harmonic driver) adds the even harmonics. Thus, breaking of the space-time symmetry softens the selection rules. The same situation occurs in our experiment upon changing the polarization of one or both colors to elliptical [by rotating the waveplates away from $(\alpha,\beta)=(45^0,45^0)$ ]. This would break the 2.95-fold symmetry (see the Lissajous curves for $\alpha \neq 45^0$ or $\beta \neq 45^0$ in Fig. 1), and it is found that additional harmonics are obtained, that could be written in the form

$$\Omega = (2.95 n_1 \pm 1, \pm 3, \pm 5, ...) \omega \qquad (S3)$$

which is nothing but another way of writing Eq. 1. By controlling the ellipticity of the two pulses, the direction and impact parameter of the returning electronic trajectories can be controlled, as well as sub-optical-cycle synchronization between the subsequesnt recollision events. This directly translates into the polarization state of the emitted high harmonics. And is the reason for the high controllability over the polarization achieved in our experiment.

The reason for the high efficiency of our scheme lies in the relatively short time (compared to monochromatic HHG) that the electron spends in the continuum, hence its lateral spreading is reduced, giving rise to enhanced recombination probability. This is clearly seen in Fig. S2 which compares the HHG spectra for the counter-rotating case [$(\alpha,\beta)=(45^0,45^0)$] and for the case where the two drivers are collinearly-polarized .

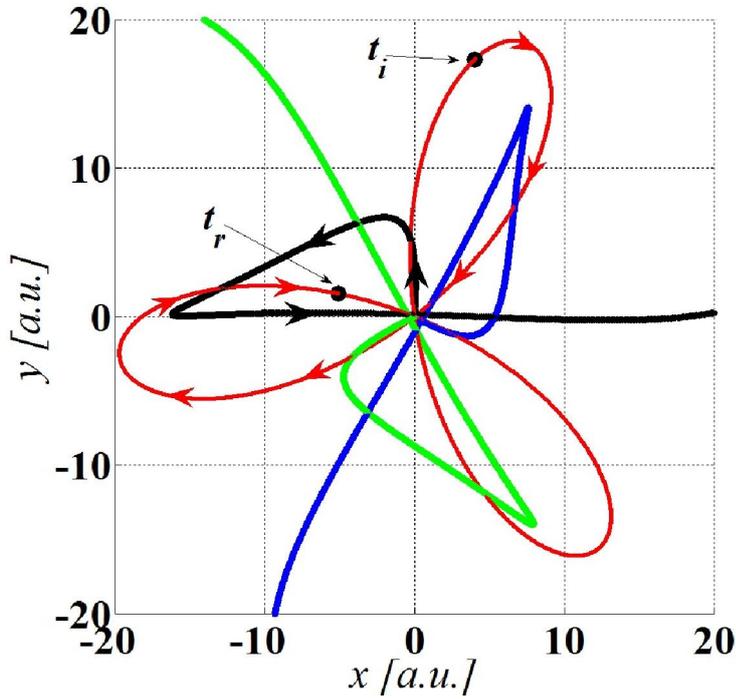

Figure S.1: electron cut-off trajectories for the counter-rotating circular drivers case, presented in a coordinate system for which one of the recollisions (black solid curve) occurs along the x-axis. The birth and recombination times of that recollision are labeled as $t_i$ and $t_r$, respectively on the Lissajous curve (solid red). The electron is being born in the positive y-direction at time $t_i$, changes direction twice and recollides with the ion with an almost head-on recollision at time $t_r$ along the positive x-direction. Due to the 2.95-fold rotational symmetry in the shape of the electric field, almost one third of an optical cycle later a similar recollision occurs (green solid curve), but along a direction which forms an angle of $2\pi/2.95$ with the positive axis. The blue solid curve is again an otherwise identical recollision, which is born one-third of an optical cycle after the former one, from a direction which forms an $4\pi/2.95$ angle with the positive axis.

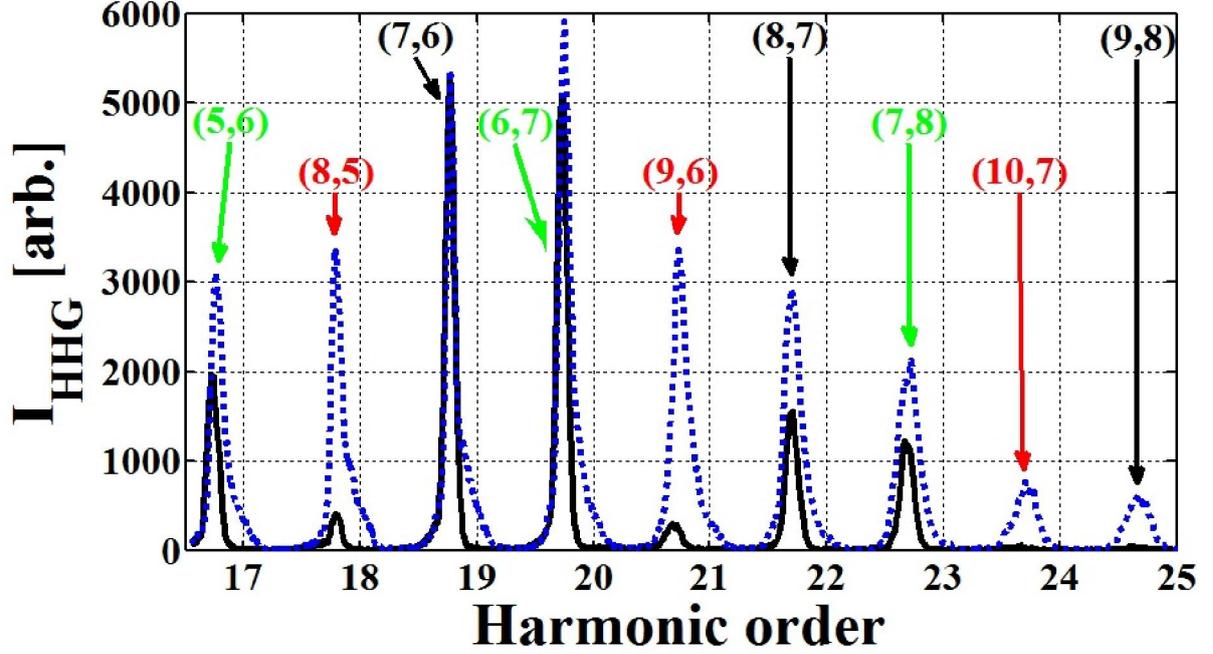

Figure S.2: HHG spectra for bi-chromatic drivers with counter-rotating circular polarization [(α,β)=(45⁰,45⁰)] that results with harmonics with circular polarization (solid black line) and for the same drivers with the only difference that they are collinearly-polarized and produce linearly-polarized harmonics (dashed blue line). The channels are indicated. Clearly, the generation efficiency of circularly-polarized and linearly-polarized harmonics is similar.

**2. Strong-field corrections**

The relation between a beam's ellipticity $\varepsilon$ and spin $\sigma$ is derived by simple transformation of bases. Any general elliptically-polarized field (written for simplicity in the orientation ellipse's principle axes $\mathbf{e}_f, \mathbf{e}_s$) having ellipticity $\varepsilon$ (taken as $0 < \varepsilon < 1$) could be written as a superposition of two fields with counter-rotating circular polarization:

$$E(t) = \sqrt{\frac{2}{1+\varepsilon^2}} E_0 \cos(\omega t + \varphi) \mathbf{e}_f + h\sqrt{\frac{2\varepsilon^2}{1+\varepsilon^2}} E_0 \sin(\omega t + \varphi) \mathbf{e}_s$$

$$= \sqrt{\frac{(1+\varepsilon)^2}{2+2\varepsilon^2}} E_0 \left[\cos(\omega t + \varphi)\mathbf{e}_f + h \cdot \sin(\omega t + \varphi)\mathbf{e}_s\right] + \sqrt{\frac{(1-\varepsilon)^2}{2+2\varepsilon^2}} E_0 \left[\cos(\omega t + \varphi)\mathbf{e}_f - h \cdot \sin(\omega t + \varphi)\mathbf{e}_s\right]$$

where $E_0$ is some amplitude, $\varphi$ is a general phase and $h = \pm 1$ is the helicity. In a ket notation

$$|E\rangle = \sqrt{\frac{(1+\varepsilon)^2}{2+2\varepsilon^2}} E_0 |+h\rangle + \sqrt{\frac{(1-\varepsilon)^2}{2+2\varepsilon^2}} E_0 |-h\rangle \qquad (S4)$$

where the states $|+1\rangle$ and $|-1\rangle$ correspond to left- and right-circular polarizations, respectively. The spin of these basis states is $\hbar$ and $-\hbar$, respectively: $\hat{L}_z|h\rangle = \hbar h|h\rangle$. The expectation value of the field's spin $\sigma = \dfrac{\langle E|\hat{L}_z|E\rangle}{\langle E|E\rangle}$ is therefore

$$\sigma = \hbar h \left[ \frac{\left|\sqrt{\dfrac{(1+\varepsilon)^2}{2+2\varepsilon^2}}E_0\right|^2 - \left|\sqrt{\dfrac{(1-\varepsilon)^2}{2+2\varepsilon^2}}E_0\right|^2}{\left|\sqrt{\dfrac{(1+\varepsilon)^2}{2+2\varepsilon^2}}E_0\right|^2 + \left|\sqrt{\dfrac{(1-\varepsilon)^2}{2+2\varepsilon^2}}E_0\right|^2} \right] = \hbar \frac{2h\varepsilon}{1+\varepsilon^2} \qquad (S5)$$

Moreover, if the field $E(t)$ is obtained by passing a linearly-polarized field through a half-waveplate whose fast axis forms an angle $\alpha$ with the polarization direction of the incoming linear light, we have $\varepsilon = \tan\alpha$, $h = 1$ and get $\sigma/\hbar = 2\varepsilon/(\varepsilon^2+1) = \sin(2\alpha)$. This relation is used to represent the spin of the 800nm beam and a similar expression with a waveplate angle $\beta$ represents the spin of the 410nm beam.

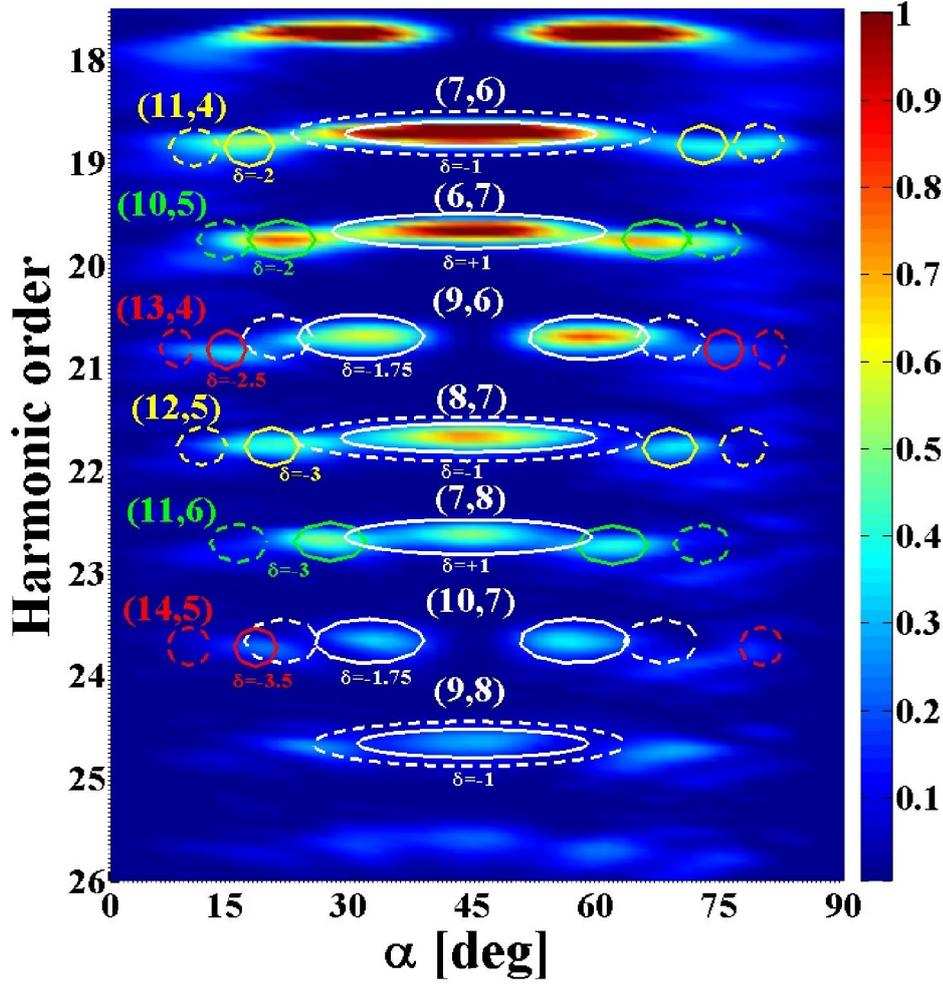

Figure S.3: same as Figure 2c. Ellipses: "existence regimes" for harmonic channels ($n_1,n_2$) with (solid lines) and without (dashed lines) strong-field correction $\delta_{(n_1,n_2)}$ which enters the conservation law (Eq. 2). While qualitatively matching, additional spin needs to be added in order for the numerical "existence regimes" and the ones predicted by the spin conservation requirement to quantitatively agree.

### 3. Polarization analysis of high harmonics

Suppose the polarization ellipse of some harmonic in the lab coordinate system is $\bar{\mathbf{E}}_0(\Omega) = \bar{E}_x(\Omega)\mathbf{e_x} + \bar{E}_y(\Omega)\mathbf{e_y}$. We write this field in a vector form as

$$\begin{pmatrix} \bar{E}_x \\ \bar{E}_y \end{pmatrix}$$

The harmonic field passes the HHG polarizer, whose main transmission axis makes an angle $\theta$ with respect to $\mathbf{e_x}$, whose Fresnel coefficients for transmission are $r_s$, $r_p$ (giving extinction ofroughly $|r_s|^2/|r_p|^2 \cong 40:1$). Later on, the field passes the HHG spectrometer, whose principal axes are $\bar{\mathbf{E}}_0(\Omega) = \bar{E}_x(\Omega)\mathbf{e_x} + \bar{E}_y(\Omega)\mathbf{e_y}$, which acts as an additional polarizer that favors light polarized along $\mathbf{e_y}$ by a factor of ~2.5:1. Written as a product of transmission matrices and switching back and forth between the principal axes of the polarizer and the lab coordinate system, the final field hitting the CCD is given by

$$\begin{pmatrix} \bar{E}_{1x} \\ \bar{E}_{1y} \end{pmatrix} = \begin{pmatrix} r'_s & 0 \\ 0 & r'_p \end{pmatrix} \begin{pmatrix} \cos\theta & -\sin\theta \\ \sin\theta & \cos\theta \end{pmatrix} \begin{pmatrix} r_s & 0 \\ 0 & r_p \end{pmatrix} \begin{pmatrix} \cos\theta & \sin\theta \\ -\sin\theta & \cos\theta \end{pmatrix} \begin{pmatrix} \bar{E}_x \\ \bar{E}_y \end{pmatrix} \quad \text{(S6)}$$

That is, the intensity is $I(\Omega) = |\bar{E}_{1x}(\Omega)|^2 + |\bar{E}_{1y}(\Omega)|^2$. Since the field $\bar{\mathbf{E}}_0(\Omega) = |\bar{E}_x(\Omega)|e^{i\phi_x(\Omega)}\mathbf{e_x} + |\bar{E}_y(\Omega)|e^{i\phi_y(\Omega)}\mathbf{e_y}$ is complex, 4 variables are present. The input parameters are $I_{(n_1,n_2)}(\theta), r_s, r_p, r'_s, r'_p$ where $I_{(n_1,n_2)}(\theta) \equiv I(\Omega = n_1\omega_1 + n_2\omega_2, \theta)$ is the intensity measured for harmonic channel $(n_1, n_2)$, when the HHG polarizer is set at angle $\theta$. By rotating this polarizer we obtain an intensity scan $I_{(n_1,n_2)}(\theta)$ as the one shown in Figure S.4.

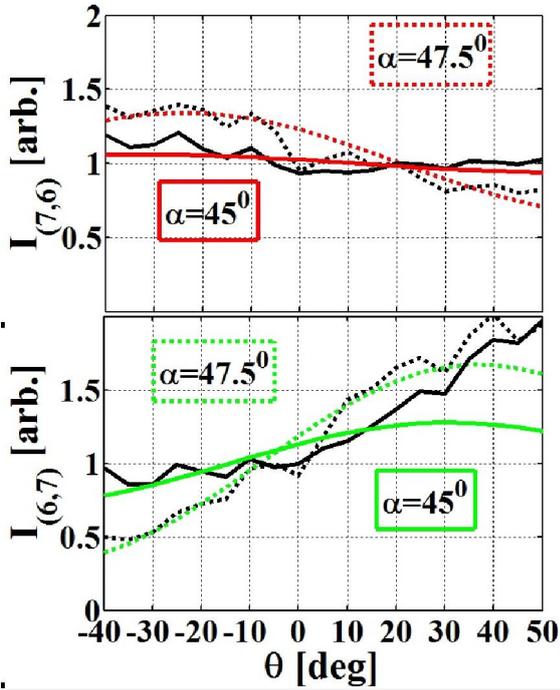

Figure S.4: Polarization traces from which the experimental ellipticites in Fig.3b are derived. Upper panel: harmonic spectra of channel (7,6) taken for (α,β)=(45⁰,45⁰) (solid black line) and fitted to a Malus-law type function (solid red line), and measured and fitted spectra for (α,β)=(47.5⁰,45⁰) (dashed black and red lines, respectively). Lower panel: same as upper panel, but for channel (6,7). Black lines: measured spectra, green lines: fitted spectra.

By fitting procedure we find the 4 functions $|\bar{E}_x|, |\bar{E}_y|, \phi_x, \phi_y$ from which the orientation angle of the harmonic polarization ellipse and the ellipticity are obtained

$$tg(2\phi) = \frac{2|\overline{E}_x||\overline{E}_y|}{|\overline{E}_x|^2 - |\overline{E}_y|^2}\cos(\phi_y - \phi_x)$$

$$\varepsilon^2 = \frac{|\overline{E}_x|^2 + |\overline{E}_y|^2 - \sqrt{\left(|\overline{E}_x|^2 - |\overline{E}_y|^2\right)^2 + 4|\overline{E}_x|^2|\overline{E}_y|^2\cos^2(\phi_y - \phi_x)}}{|\overline{E}_x|^2 + |\overline{E}_y|^2 + \sqrt{\left(|\overline{E}_x|^2 - |\overline{E}_y|^2\right)^2 + 4|\overline{E}_x|^2|\overline{E}_y|^2\cos^2(\phi_y - \phi_x)}}$$

(S7)

### 4. Photon pairing

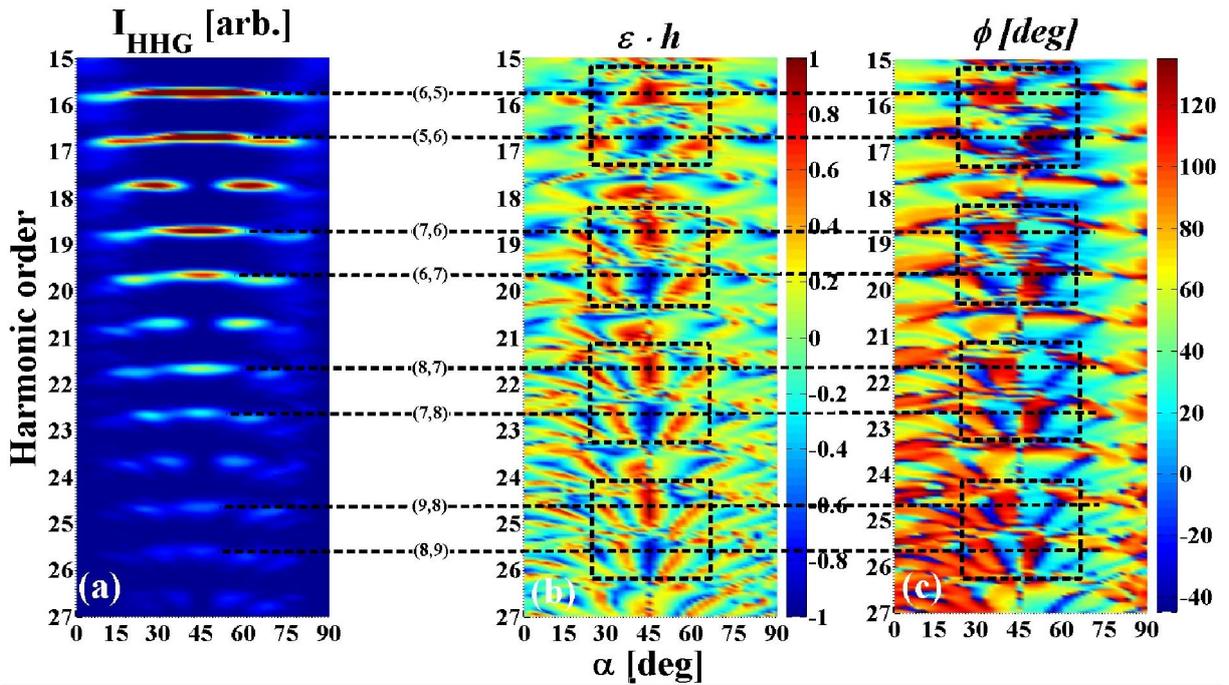

Figure S.5: Harmonic spectra (a), product of ellipticity times helicity (b) and orientation angle of the polarization ellipse (c) for the α–scan experiment. It is clearly seen that harmonics belonging to the group $(n_1, n_1-1)$ have similar α–dependences of both $\varepsilon \cdot h$ and $\phi$. Same applies to the channels of the form $(n_1, n_1+1)$. Moreover, the $(n_1, n_1-1)$ and $(n_1, n_1+1)$ groups of channels exhibit mirror-like polarization dynamics.